\documentclass[12pt]{article}
\usepackage{graphicx}
\graphicspath{{img/}}
\usepackage{subcaption}
\usepackage{amsmath}
\usepackage{amssymb}
\usepackage{csquotes} 
\usepackage{physics}
\usepackage{slashed}

\usepackage{mathtools}
\usepackage{amsfonts}
\numberwithin{equation}{section}

\usepackage{xcolor}
\usepackage{hyperref}
\usepackage{placeins} 

\newcommand{\beq}{\begin{equation}}   
	\newcommand{\eeq}{\end{equation}}
\newcommand{\beqn}{\begin{eqnarray}}   
	\newcommand{\eeqn}{\end{eqnarray}}

\newcommand{\pt}{\partial}

\newcommand{\Zc}{{\mathcal Z}}
\newcommand{\Wc}{{\mathcal W}}
\newcommand{\Lc}{{\mathcal L}}
\newcommand{\Fc}{{\mathcal F}}
\newcommand{\Dc}{{\mathcal D}}
\newcommand{\phiv}{\varphi}

\def\none{${\mathcal N}=1\;$}

\newcommand{\gsim}{\lower.7ex\hbox{$
\;\stackrel{\textstyle>}{\sim}\;$}}
\newcommand{\lsim}{\lower.7ex\hbox{$
\;\stackrel{\textstyle<}{\sim}\;$}}

\begin{document}

\hypersetup{%
	linkbordercolor=blue,
}

\begin{titlepage}

\begin{flushright}
FTPI-MINN-26-04 \\ UMN-TH-4520/26 
\end{flushright}


\begin{center}
	\Large{{\bf \boldmath
		Degenerate vortices and world-line instantons \\ in three-dimensional gauge theories
	}}

\vspace{5mm}
	
{\large  \bf Evgenii Ievlev and Mikhail Shifman}

\end{center}
\begin{center}
{\it  Department of Physics,
University of Minnesota,
Minneapolis, MN 55455}\\[5pt]
{\it  William I. Fine Theoretical Physics Institute,
University of Minnesota,
Minneapolis, MN 55455}\\
\end{center}

\vspace{5mm}

\begin{center}
{\large\bf Abstract}
\end{center}

In this paper we continue the study of particle-like topological solitons with degenerate masses and their mixing due to world line instantons. Previously, this phenomenon was studied in 1+1-dimensional setups. Here we take a step further and consider degenerate vortices in 2+1 dimensions. We find that, while classically such vortices may be degenerate, they generally mix and split at the quantum level. Supersymmetry protects BPS-saturated vortices only when the number of supercharges in the bulk is large enough.

\end{titlepage}

{
\setcounter{tocdepth}{2}
\tableofcontents
}

\section{Introduction}
\label{intro}

Whenever a theory has multiple vacua that are degenerate in energy at the classical level, it typically also admits quasiclassical field configurations of nontrivial topology, interpolating between these vacua. 
A familiar example comes from $(0+1)$-dimensional quantum mechanics (QM) with a double-well potential: in the purely bosonic system, instantons generate a nonzero overlap between the two perturbative ground states, while in the supersymmetric counterpart the tunneling amplitude is suppressed.
If the tunneling is not suppressed, the would-be degenerate ground states become split in energy.

In higher dimensions, it becomes possible to have multiple solitons (kinks, etc.) that are degenerate in energy.
As such, they may mix on the quantum level --- an effect that is known in condensed matter
\cite{Takagi:96a,DubeStamp1998,ShibataTakagi2000}, and that recently gained attention in the particle physics community \cite{Evslin:2025zcb,Ievlev:2025ibm}.

While the previous field-theoretic studies focused on 1+1D theories, here we take a step further and consider the 2+1D setup.
In this case, one can consider codimension-1 and codimension-2 solitons.
The former are the domain lines (walls); however, they are not the main focus of this paper.
Here, we concentrate on vortices, which fall into the latter category above.

We cover several examples where the vortices have internal features.
This can occur in both Abelian and non-Abelian gauge theories.
Classically, the presence of the internal structure (like discrete moduli) leads to having several species of vortices with the same mass and topological charge.
On the quantum level, the mass degeneracy is generically lifted.
This can happen due to world line instantons, which are the instantons in the effective quantum mechanics living on the vortex world line.
As a result, the vortex degeneracy is lifted.

The instanton tunneling amplitude can be suppressed in supersymmetric quantum mechanics, e.g. in the case with a double-well potential.
Therefore, if the vortex is Bogomol'nyi-Prasad-Sommerfield (BPS) saturated in a theory with large enough supersymmetry in the bulk, we can expect that the degeneracy will be protected.
We find that with eight supercharges the vortices stay degenerate, while for lesser supersymmetry their degeneracy may be lifted.

We begin this paper by discussing some basic 3D models in Sec.~\ref{sec:models_intro}.
We focus on the cases with $\mathcal{N}=1$ supersymmetry and their bosonic counterparts.
The issue of BPS-saturated vortices is also discussed there.
After that, we turn to the main subject of this paper, namely, degenerate vortices, starting with Abelian examples in Sec.~\ref{sec:deg_vort_abelian}.
In Sec.~\ref{sec:deg_vort_non-ab} we continue with non-Abelian gauge theories, which can support vortices with large degeneracy.
Our conclusions are summarized in Sec.~\ref{sec:concl}.
Appendix~\ref{sec:3d_susy} provides some details on $\mathcal{N}=1$ supersymmetry in three spacetime dimensions.

\section{Models with \boldmath  $\mathcal{N}=1$ supersymmetry}
\label{sec:models_intro}

We are going to start here by reviewing some basic setups for three-dimensional models with and without supersymmetry.
These models (and their generalizations) will be studied in more detail below.
Additional details on three-dimensional supersymmetry are reviewed in Appendix~\ref{sec:3d_susy}.

\subsection{Scalar model and its gauging}

The minimal example is an $\mathcal{N}=1$ supersymmetric Wess-Zumino type model (two supercharges).
The Lagrangian for such a model can be written as
\beq
S=\!\frac{1}{2}\int d^3 x
\left\{
\partial_\mu\phiv^a 
\partial^\mu\phiv^a + \bar\psi^a \,i\gamma^\mu 
\partial_\mu \psi^a -\left[\partial_a{\cal W}(\phiv^{c})\right]^2 - \Big(\partial_a \partial_b {\cal W}(\phiv^c)\Big)
\bar\psi^a\psi^b \right \},
\label{sigmaLp}
\eeq
where $\varphi^a$ are real scalar fields, $a$ is a flavor index, and $\Wc (\phiv^a)$ is the superpotential depending on all the $\phiv^a$ fields.
Below we will discuss an example with
\beq
\Wc(\phiv^a) =\frac{m}2 (\phiv^a\phiv^a) -\frac{g}{8} (\phiv^a\phiv^a) ^2 \,.
\label{rsp}
\eeq
Finally, $\bar\psi= \psi^\dagger \gamma^0 = \psi\gamma^0$ and $\psi$ is a Majorana two-component spinor.
This model has two real supercharges.  
In 3D there is no chirality; therefore, no chiral subspace can be constructed. 

When the number of flavors in \eqref{sigmaLp} is even, one can define a $U(1)$ symmetry, which then can be gauged.
Of course, the superpotential must be invariant with respect to the symmetry transformations.
For example, in the case with two flavors, we can combine two real fields into one complex field $\phi$ and two Majorana spinors into a complex two-component spinor $\Psi$,
\beq
\phi =\frac{1}{\sqrt 2}\left(\phiv^1+ i \phiv^2\right),\qquad \Psi =\frac{1}{\sqrt 2}\left( \psi^1+i\psi^2\right) \,.
\label{complex}
\eeq
There is a natural global $U(1)$ symmetry acting on these complex fields.
Gauging this symmetry brings up the gauge field (photon) $A_\mu$ which in 3D carries one physical component, together with its superpartner --- a neutral Majorana field $\lambda$ (photino).
The Lagrangian of 3D ${\mathcal N}=1$ supersymmetric quantum electrodynamics (SQED) in the  Wess-Zumino gauge takes the form
\beqn
{\mathcal L}_{\rm 3D \,\,SQED} &=&\frac{1}{e^2}\left(-\frac 1 4 F_{\mu\nu} F_{\mu\nu}+ 
\frac{i}{2} \lambda^\alpha \partial_{\alpha\beta}\lambda^\beta \right)
\nonumber\\[2mm]
&+& ({\mathcal D}_\mu\phi)^\dagger{\mathcal D}^\mu\phi  +
i\left((\Psi^\alpha)^\dagger (\mathcal D)_{\alpha\beta}\Psi^\beta \right)
\nonumber\\[2mm]
&-& U(\phi, \phi^\dagger) + \left[ \lambda\gamma^0 \Psi \phi^\dagger  +{\rm H. c.}
\right] -\frac{i}{2} \,\frac{\partial^2 {\mathcal W}}{\partial \phi\partial \phi^\dagger}\Psi^\dagger\gamma^0 \Psi\,,
\label{35spp}
\eeqn
where
\begin{equation}
	F_{\mu\nu} =\partial_\mu A_\nu - \partial_\nu A_\mu\,,\qquad {\mathcal D}_\mu=\partial_\mu -A_\mu\,,\quad \partial_{\alpha\beta} =(\gamma^0\gamma^\mu)_{\alpha\beta}\partial_\mu\,,
\end{equation}
and the potential
\begin{equation}
	U = (\phi\,\phi^\dagger)\left[ m-g (\phi\,\phi^\dagger) \right]^2
\label{sextic_potential}
\end{equation}
follows directly from the superpotential \eqref{rsp}.
Note that this potential is sextic, which is renormalizable in 2+1D.

The theory \eqref{35spp} can be obtained from the 4D ${\cal N}=1$ SQED (four supercharges) upon dimensional reduction and discarding the fields that are absent in
3D ${\mathcal N}=1$  SQED%
\footnote{Equation (\ref{35spp}) may be compared with Ref.  \cite{Olmez:2008au} where vortices in a 3D reduction of 4D SQED with four supercharges were considered (see Eq. (7) in \cite{Olmez:2008au}  and Sect. 10.6.9 in \cite{Shifman:2012zz}).  In 4D the photino field $\lambda$ is described by a four-component Majorana spinor, while in  3D ${\mathcal N}=1$ SQED the photino is a two-component Majorana spinor. In passing from 4D to 3D with two supercharges we discard the $A_2$ component of the gauge field -- it is absent in 3D ${\mathcal N}=1$ (it becomes a  scalar field $N$ in the notation of \cite{Olmez:2008au}). Finally,  all tilded fields present in Eq. (7) of \cite{Olmez:2008au} disappear. Note that  in ${\mathcal N}=1$ 3D SQED there is no Fayet-Iliopoulos term and its $D$-term companions. It is also important to emphasize that in ${\mathcal N}=1$  SQED in three dimensions no bona fide chiral superpotential exists because there are no chiral sub-spaces. Instead,  we have a superpotential (\ref{quartsup}). Our definition of 4D $\to$  3D reduction is as follows:  out of four coordinates we drop $y$ and  each vector loses its $y$ component, see Ref. \cite{Shifman:2012zz}. In colloquial language $\Zc_\mu$ is often referred to as central charge, although it is not quite accurate because $\Zc_\mu$ does not commute with the generators of rotations. A more accurate name in this case is the {\em brane charge} suggested by N. Seiberg and Z. Komargodski.} 
or, alternatively, constructed directly in 3D; see Appendix~\ref{sec:3d_susy}. 

The model under consideration has two vacua with $\expval{\phi}$ given by:
%
\begin{subequations}
	\begin{align}
		\label{basic_vacua_sym} &\phi_{*\text{sym}} =0 	\,;	 \\
		\label{basic_vacua_as} &\phi_{*\text{as}} = v e^{i \alpha} \,, \quad v = \sqrt{\frac{m}{g}}	\,.
	\end{align}
	\label{basic_vacua}
\end{subequations}
These vacua are degenerate in energy, at least at the classical level.

The first vacuum, Eq.~\eqref{basic_vacua_sym}, is the normal, or symmetric, or non-su\-per\-con\-duc\-ting phase.
It is invariant under the $U(1)$ phase rotations.
The photon in this phase is massless, while the complex scalar has mass $m$.

The second vacuum, Eq.~\eqref{basic_vacua_as}, is the asymmetric, or superconducting, or Higgs phase.
Strictly speaking, it is a whole manifold, but all its points are gauge equivalent to each other.
The $U(1)$ symmetry is realized non-linearly; alternatively, one can say it is spontaneously broken. 
A standard redistribution of degrees of freedom occurs in this vacuum. 
The phase of the $\phi$ field will be eaten up by the photon (thus acquiring a longitudinal degree of freedom and a mass $M_{\rm ph}\sim ev$. 
The physical Higgs field with mass
\begin{equation}
	m_{|\phi|} = 2 m = 2 g v^2
\label{mass_higgs}
\end{equation}
remains the lower component of the {\em real} \, Higgs superfield.
The vacuum expectation value (VEV) $v$ is large at weak coupling, i.e., $g \ll 1$.
This paves the way to study, in addition to vortices in the Higgs phase, also domain lines separating the two phases.

In three dimensions, vortices are particles with finite masses. 
Below we will see that, in some modified versions of this theory, the Higgsed regime exhibits a number of mass-degenerate vortices.
When supersymmetry is absent, these vortices mix through instantons.
In 3D this is a new phenomenon similar to that discussed in 2D in \cite{Evslin:2025zcb,Ievlev:2025ibm}. 
Supersymmetrizing the model, one can force the instanton contribution to vanish because of the fermion zero modes. 

The model under consideration exhibits other interesting phenomena, such as the erasure of vortices crossing the domain line from the Higgsed to the un-Higgsed phase \cite{Dvali:2022rgx} and non-trivial effects associated with the inclusion of the Chern-Simons term. 
These two topics will be considered in a forthcoming publication \cite{IevlevShifmanInPrep2026}.

\subsection{Vortices in gauged 3D model, Higgs regime}

As explained in \cite{Shifman:2012zz}, Sec. 11.1.2.2, in three-dimensional \none supersymmetric QED (SQED) there is no central charge corresponding to vortices.
This implies that the vortices cannot be BPS saturated.
However, we will see later that they still can be approximately BPS with a judiciously chosen superpotential.

Recall that the model \eqref{35spp} has two vacua, one symmetric at $\phi_{*\text{sym}}=0$, and another asymmetric with $\phi_{*\text{as}}=v e^{i \alpha}$. 
Let us have a closer look at the second vacuum.

First, we will study the possibility of the Bogomol'ny completion for the vortex equation. 
For static field configurations, the energy functional takes the form (in the gauge $A_0=0$):
\beq
{\mathcal E}[\vec{A} (\vec x), \phi(\vec x)] = \int dx_1 dx_2\left[\frac{1}{4e^2}
F_{ij}F_{ij}+|
{\mathcal D}_i\phi |^2+ U(\phi)
\right] \,,
\eeq
where $U(\phi)$ is presented in Eq. (\ref{sextic_potential}).
It is rather obvious that with this superpotential the BPS completion is impossible. 

Thus, to find the vortex solution we have to solve a second-order differential equation.
The relevant ansatz and the boundary conditions are the same as in Sec. 3.1.3 of \cite{Shifman:2012zz} (here, $(r,\alpha)$ are the polar coordinates on the 2D space):
\begin{equation}
	\phi(r,\alpha) = v f_\phi( r ) e^{i \alpha} \,, \quad
	A_i(r,\alpha) = - \varepsilon_{ij} \frac{x_j}{ r^2 }[1 - f_A( r )] \,;
\label{vortex_ansatz_simple}
\end{equation}
\begin{equation}
	\begin{aligned}
		f_\phi(0) &= 0 \,, \quad f_\phi(\infty) &= 1 \,, \\
		f_A(0) &= 1 \,, \quad f_A(\infty) &= 0 \,.
	\end{aligned}
\label{vortex_boundary_conditions}
\end{equation}
However, the profile functions are different from the BPS case.

Since the BPS completion is impossible, both supercharges are broken on the vortex solution.
Then on such a vortex we will have two fermion zero modes.
Can we change the superpotential in such a way that the BPS saturation becomes possible, at least at the classical level?
The answer is yes, provided we add to the theory one extra scalar field $h$, neutral under the gauge $U(1)$, and choose 
the superpotential as follows,
\beq
\Wc_h = g h\Big(\Phi^\dagger\Phi - v^2\Big)\,,
\label{W_almost_bps}
\eeq
implying
\beq
U= \frac{1}{2} g^2\Big( v^2-\phi^\dagger\phi\Big)^2 +g^2h^2 \phi^\dagger\phi
\label{55s}
\eeq
with the ground state at $|\varphi_{\rm vac}| = v$, $h=0$ and dim\,$g =[m^{1/2}]$.
Then in the classical solution we can put  $h=0$, arriving at basically the same classical first-order equations in the bosonic sector as in Sec. 11.5.1
of \cite{Shifman:2012zz}\footnote{The emergence of the term $g^2\Big( v^2-\varphi^\dagger\varphi\Big)^2$ in the potential in Sec. 11.5.1 of \cite{Shifman:2012zz} is due to the Fayet-Iliopoulos $\xi$ term. In our case it cannot be introduced, we have 
to generate it from $F$ terms.}
(cf. also  Sec. 3.1.3 there),
\begin{equation}
	\begin{aligned}
		B - g^2 (|\phi|^2 - v^2) &= 0 \,, \\
		(\mathcal{D}_x + \mathcal{D}_z) \phi &= 0 \,.
	\end{aligned}
\end{equation}
The bosonic profile solution is also the same as in the BPS case, only supplemented by $h=0$ for the extra scalar field that we introduced here.
The difference from the BPS case is going to show up in the zero modes and quantum corrections.

From the action (\ref{35spp}) we obtain a pair of coupled equations of motion for the fermion fields,
\beqn
&& i\partial_{\alpha\beta}\lambda^\beta +e^2 \big(\Psi^\dagger +\Psi\big)_\alpha =0\,,
\nonumber\\[2mm]
&&i{\mathcal D}_{\alpha\beta}\Psi^\beta +\phi \lambda_\alpha + \frac{\partial^2 {\mathcal W}}{\partial \phi\partial \phi^\dagger}\Psi_\alpha =0\,.
\eeqn
On the vortex background, these equations describe the fermion zero modes.

To conclude this section, let us note that the theory described here has a domain line (wall) solution separating the two phases \eqref{basic_vacua}; see Appendix~\ref{sec:DL_charge}.
It would be interesting to study the degenerate ``domain wall'' vortices, which would be the analogs of the objects studied in \cite{Dumitrescu:2025fme} in the context of 4D gauge theories.

\section{Degenerate vortices inside an Abelian superconductor}
\label{sec:deg_vort_abelian}

In this section we come to the main topic of this study, namely, degenerate vortices.
We are going to start with the case of the $U(1)$ gauge theory.
Vortices inside a non-Abelian superconductor will be described in the next section.

\subsection{$\mathbb{Z}_2$ vortices}
\label{sec:Z2_vort}

Here we extend the analysis presented in \cite{Ievlev:2025ibm} and study  another example -- perhaps more realistic for a condensed matter system.
This example is motivated by the study of degenerate vortices in gauged sigma models \cite{Alonso-Izquierdo:2014cza,Alonso-Izquierdo:2016ieq}.
In the setup of \cite{Alonso-Izquierdo:2014cza,Alonso-Izquierdo:2016ieq}, the sigma model target space imposes a rigid constraint on the fields, and world line instantons are {\em not possible}.
We will relax the rigid constraint, introducing a penalizing term in the potential, which will make the instanton-induced tunneling possible. 
A similar model was studied in \cite{Adhikari:2017oxb} (in a somewhat different notation) and  the vortex solutions were constructed.
Here, we go one step further and discuss a possible mixing of these vortices.

\subsubsection{The model}

We consider a 2+1D Abelian  theory with the U(1) gauge group, a complex scalar field $\phi$ of charge 1, and a real scalar $\chi$ (charge 0).
The Lagrangian ${\cal L}$ consists of two parts \cite{Shifman:2012vv,Shifman:2013oia,Monin:2013kza},
\beq
{\cal L} = {\cal L}_{\rm QED} +{\cal L}_\chi \,.
\label{twosum}
\eeq
The first piece is the standard scalar QED, for now without SUSY:
\beqn
\label{tpi15}
{\cal L}_{\rm QED} &=& -\frac{1}{4e^2}F_{\mu\nu}^2 + \left| {\mathcal D}^\mu\phi\right|^2 -V(\phi) \,,\\[2mm] 
V&=& e^2 (|\phi |^2 -v^2)^2\,,
\label{tpi16}
\eeqn
where 
\beq
F_{\mu\nu} = \partial_\mu A_\nu -\partial_\nu A_\mu\,,\qquad {\mathcal D}_\mu\phi = (\partial_\mu -iA_\mu )\phi \,.
\eeq
The second piece describes an extra real scalar field $\chi$ coupled to the QED sector in a special way,
\beqn
{\cal L}_\chi &=& \frac{1}{2}\partial_\mu \chi \, \partial^\mu \chi - U(\chi, \phi)\,,
\label{14s}\\[2mm]
U &=&  \gamma\left[\left(-\mu +|\phi |^2
\right)\chi^2 + \beta  \chi^4 \right],
\label{15s}
\eeqn
where $\chi$ is a real scalar field, 
$\gamma$ is a (positive) coupling constant, and 
for simplicity we will assume  $\beta \gsim 1$.
The parameters $\mu$ and $v$ are also real and positive.
In order to obtain desired degenerate vortices, we impose the following constraints\footnote{Couplings in the inequalities \eqref{12mv} should not be vastly different, however; a factor of 2 or so between them is acceptable. This will guarantee that $m_\chi \sim m_\phi$, cf. Eq.~\eqref{36s_mass}.}
on the couplings:
\beq
\mu \lesssim v^2\,,\qquad e^2 \lesssim \gamma \,.
\label{12mv}
\eeq
Note the similarity of the two potentials in Eq.~\eqref{55s} on one hand, and in Eqs.~\eqref{tpi16} and \eqref{15s} on the other hand.

The nonvanishing vacuum expectation value of $\phi$, $|\phi|_{\rm vac} \neq 0$, Higgses the gauge field. In the vacuum 
\beq
|\phi|_{\rm vac}^2 = v^2\,,\qquad \chi =0 \,.
\label{Z2_abelian_vac}
\eeq
The Lagrangian  (\ref{twosum}) has a $\mathbb{Z}_2$ symmetry acting as
\begin{equation}
	\chi \to - \chi \,,
\label{Z2_transformation}
\end{equation}
and it is unbroken in the vacuum \eqref{Z2_abelian_vac}.
The expectation value of $\phi$ and  Eq.  (\ref{12mv}) imply that
$\chi $ is stable, no vacuum condensate of $\chi$ develops, and the global $\mathbb{Z}_2$ symmetry remains unbroken. The masses of the photon, physical Higgs, and the neutral scalar boson $\chi$  are 
\beq
m_A=m_\phi = \sqrt{2}\, ev\,, \qquad m_\chi = \sqrt{2\gamma}\,  \sqrt{v^2-\mu}\,.
\label{36s_mass}
\eeq

The minimal winding-one vortex in this theory has the following structure.
Asymptotically far from the vortex core, we have
\begin{equation}
	\phi = v e^{i \alpha} \,, \quad
	\chi = 0 \,, \quad
	B = 0\,.
\end{equation}
where $B = F_{12}$ is the magnetic field.
Here, $\alpha$ is the polar angle on the spatial plane, while $B$ is the magnetic field.
Inside the vortex core, however, 
\begin{equation}
	\phi \approx 0 \,, \quad
	B \neq 0 \,.
\end{equation}
From the potential \eqref{15s} one can see that the mass term for $\chi$ turns negative at the vortex core.
Therefore, $\chi\neq 0$ and is double-valued there.

Thus, one can explicitly see the appearance of a degeneracy: inside the vortex, the field $\chi$ is determined only up to a sign.
This means that a vortex spontaneously breaks the $\mathbb{Z}_2$ symmetry \eqref{Z2_transformation}. 
Now recall that the vortices in 2+1D, being localized in space, are particle-like states (in much the same way as kinks in 2D, see \cite{Ievlev:2025ibm}).
Therefore, they may or may not tunnel into each other by virtue of quantum-mechanical instantons on the vortex world lines. 
The mixing depends on whether the instanton action is finite or infinite.

Let us study the vortex properties in more detail.
For a short while let us forget about the Lagrangian ${\cal  L_\chi}$ in (\ref{14s}), i.e. we set $\gamma = 0$.
To obtain a (minimal) vortex one must find the field $\phi$ at large distances from the origin, $\phi\to ve^{i\alpha}$, where $\alpha$ is the polar angle. 
This implies in turn that at the origin $\phi$ must vanish. 
The solution to the corresponding equations of motion is discussed in detail in \cite{Shifman:2012zz}. 
In fact, the solution turns out BPS saturated, $\phi =\phi_{\rm crit \,vort}$, with the unit value of the magnetic flux,  so our ``pre-vortex''  mass can be found without calculations, just from the boundary conditions,
\beq
M_{\rm vort\,0} = 2\pi v^2\,.
\label{37s_mass}
\eeq

Now, let us reinstate the $\chi$ field and consider the coupling $\gamma > 0$.
Since the $\phi$ field still has to vanish at the vortex center, $\chi$ has a non-zero VEV there, as was mentioned above.
This VEV can be found from the potential \eqref{15s}:
\beq
(\chi^2)_{\rm vort\,core } \approx \frac{\mu}{2\beta}\,,\qquad (\chi)_{\rm vort \,core} \approx \pm\sqrt{ \frac{\mu}{2\beta}}\,.
\label{19s}
\eeq
and an ``effective mass'' inside the core is
\begin{equation}
	m_{\chi,{\rm vort\,core }} \sim \sqrt{\mu\gamma} \,.
\end{equation}
We have enough parameters to guarantee that the latter is
close to $m_\chi$ in Eq. (\ref{36s_mass}) which, in turn, is close to $m_A=m_\phi$.

To estimate the actual vortex mass (with the field $\chi$ included), observe that  the expectation value of $\chi$ in the core, see Eq. (\ref{19s}), diminishes the pre-vortex mass (\ref{37s_mass}) by some fraction of $2\pi v^2$, namely,
\beq
M_{\rm vort\,0} \to M_{\rm vort} =M_{\rm vort \,0} (1-\kappa) \,,\qquad 0<\kappa <1\,,
\label{40s}
\eeq 
where $\kappa$ is a dimensionless parameter which is smaller than 1, but not significantly smaller.
In this estimate we take into account all constraints on various parameters introduced above, so that the masses $m_A=m_\phi\sim m_\chi$, which in turn implies that the size of the vortex core
\beq
R\sim m^{-1}_\phi \,.
\eeq

\begin{figure}[t]
	\centering
	\includegraphics[width=0.8\textwidth]{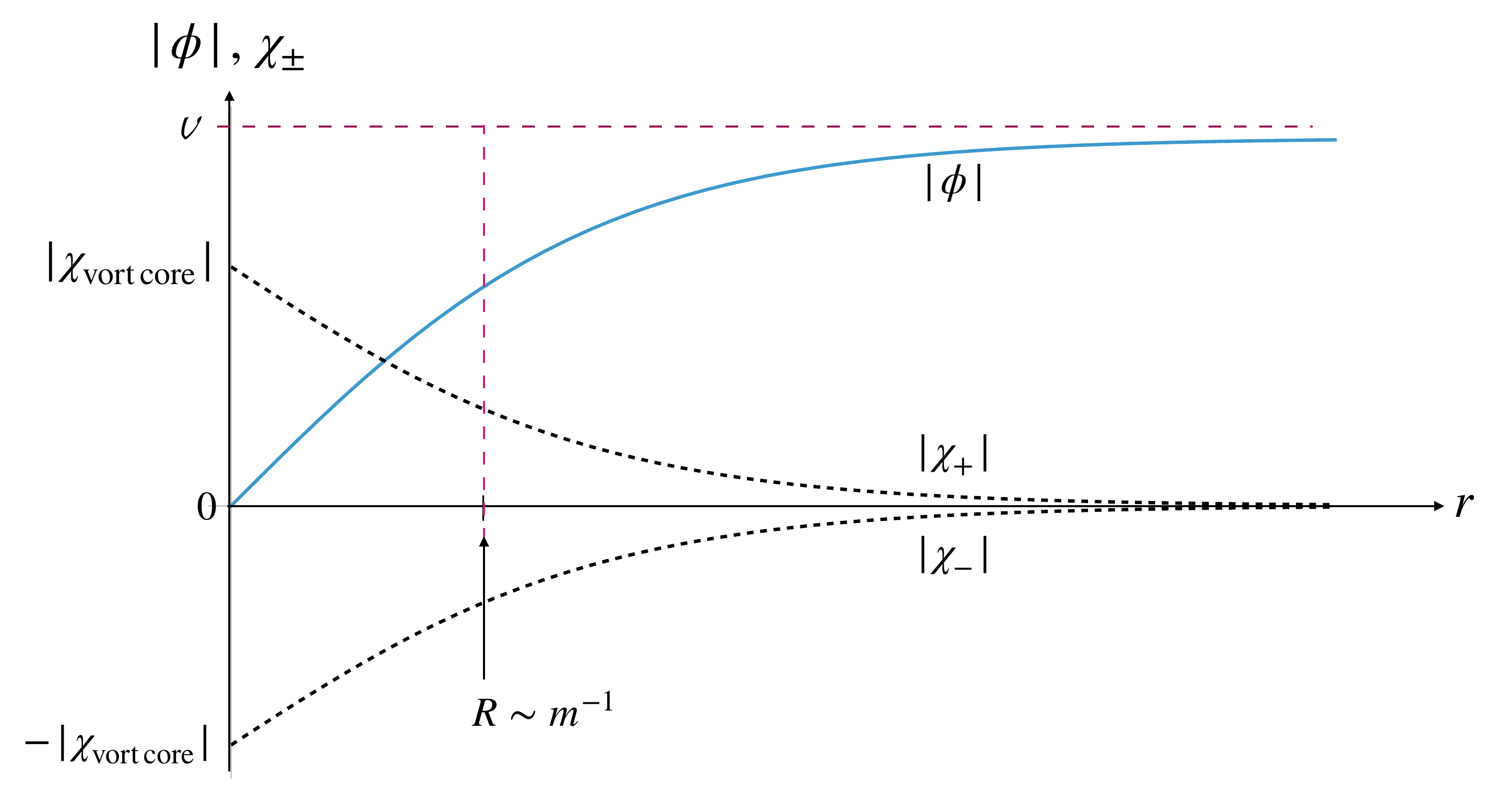}
	\caption{\small
		Schematic vortex profile functions in the model (\ref{twosum}). 
		Here $r$ is the radial coordinate, $R$ is the size of the vortex core. 
		The existence of two profile functions, $\chi_\pm$, demonstrates the spontaneous $\mathbb{Z}_2$ symmetry breaking on the vortices.
	}
	\label{figprof}
\end{figure}

\subsubsection{Tunneling between degenerate vortices in Euclidean time}

In 3D models, the vortices of the Abrikosov type can be viewed as particles.
From the standpoint of topology, they are ground states in the sector with the unit winding of $\phi$ which is  disconnected from sectors with other windings, for instance, from the topologically trivial sector, or sectors with windings $2,3$, etc.

Recall that, in our model, the global $\mathbb{Z}_2$ symmetry \eqref{Z2_transformation} is spontaneously broken in the vortex core, and $(\chi)_{\rm vort \,core}$ is double-valued (see (\ref{19s})).
Therefore, even if we arrange the topology to be the winding-1 sector, two distinct degenerate vortices coexist in this model. 
Let us call them $V_+$ and $V_-$. 
Classically, they are exactly degenerate because they are related to each other by the $\mathbb{Z}_2$ symmetry. 

Now we will discuss whether $V_+$ and $V_-$ can mix with each other quantum-mechanically. 
The mixing is realized through instantons, interpolating (in the Euclidean time $\tau$) between 
\beq
\{\phi,\,\chi\}_{\rm inst} = \left\{ \begin{array}{l}
	V_- \,\, {\rm at}\,\, \tau\to -\infty\\[2mm]
	V_+ \,\, {\rm at}\,\, \tau\to \infty
\end{array}
\right.
\eeq
If the instanton action is finite, $S_{\rm inst}<\infty$, the tunneling obviously does happen in the model under consideration. 

The instanton profile can of course be found numerically. 
We leave this exercise for  future work, limiting ourselves to some quantitative arguments.

First, the field profiles on the instanton depend on the Euclidean time $\tau$.
As $\tau$ evolves from $-\infty$ to $+\infty$, the instanton profile must continuously evolve from $V_-$ to $V_+$ without changing the winding of $\phi$ and the corresponding asymptotic behavior of the gauge field.

The potentials $V$ and $U$ in Eqs. (\ref{tpi16}) and (\ref{15s}) are nonsingular. 
The evolution of $\phi$ is expected to be small and, in any case, the asymptotic behavior of $\phi$ and $A_\mu$ stays the same. 
Therefore, the dominant contribution in the instanton action $S_{\rm inst}$ comes from the evolution of $\chi$. 

In this evolution, the energy functional reaches its maximum at $\chi=0$. 
The corresponding field configuration was referred to as the ``pre-vortex'' above. 
Now we see that the pre-vortex is unstable and is nothing other than the sphaleron. 
Its energy is given in Eq. (\ref{37s_mass}). 
Thus, the barrier separating $V_-$ and $V_+$ has the height $\kappa\,M_{\rm vort\,0} $, cf. Eq. (\ref{40s}). 
The instanton is localized in the Euclidean time $\tau$ with the characteristic size of $\Delta \tau\sim m^{-1}$. 
As a result,
\beq
S_{\rm inst}\sim \Delta\tau \kappa\,M_{\rm vort\,0} \sim 2\pi  \frac{m_\phi}{e^2}.
\label{43s}
\eeq
Below we will provide a method for deriving  this formula.

The quasiclassical approximation we rely on is valid under the condition 
\beq
\frac{v}{e}\gg 1\,.
\label{44s}
\eeq
Under this assumption, the instanton action \eqref{43s} is large, and the tunneling amplitude is in the controlled regime.

\begin{figure}[t]
	\centering
	\includegraphics[width=0.8\textwidth]{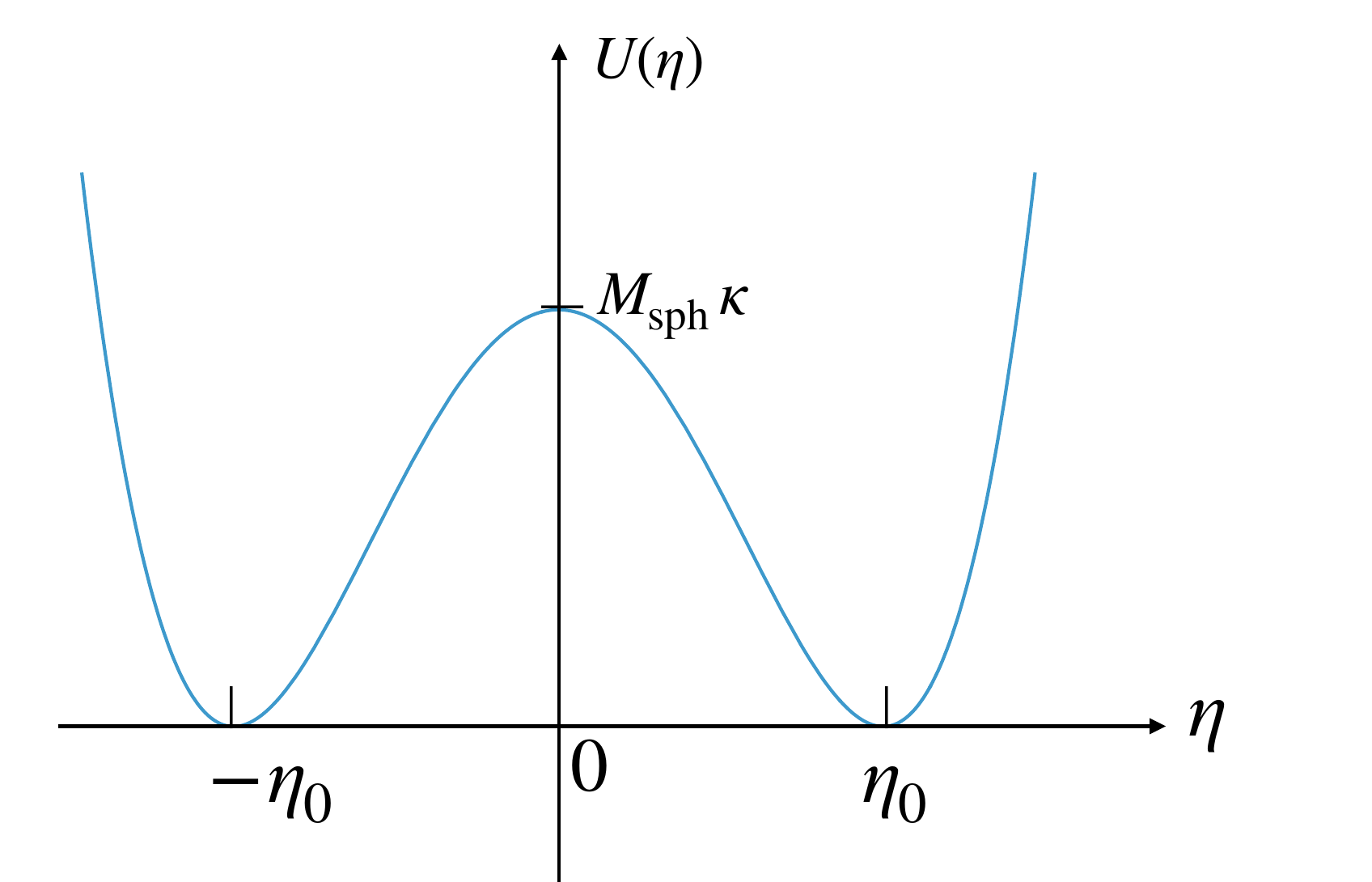}
	\caption{\small
		Effective quantum mechanics for the instanton tunneling from $V_-$ to $V_+$, see Eqs. (\ref{45s}) and (\ref{47s}).
	}
	\label{eqm}
\end{figure}

\subsubsection{Effective quantum mechanics for vortex tunneling}

To illustrate how Eq. (\ref{43s}) can be derived, let us consider an effective quantum-mechanical model that describes the $\tau$ evolution from $-\infty$ to $+\infty$. 
This quantum mechanics is supported on the world line of the vortex.

We will represent each interpolation trajectory (field profile) by a certain $\tau$-dependent parameter; for this purpose, we can choose the value of the $\chi$ field at the vortex center, $\chi(r=0)$.
Let us denote
\beqn
\chi(r=0) \!\!&\equiv&\!\! \eta\,,\\[1mm]
\eta (\tau\to -\infty)\!\! &= &\!\!-(\mu/2\beta)^{1/2},\quad \eta (\tau\to \infty) =(\mu/2\beta)^{1/2},
\eeqn 
see Eq. (\ref{19s}).
The model we suggest is the standard two-hump potential 
\beq
{\mathcal H} = \frac{1}{2}
\frac{\partial^2}{\partial\eta^2} +\lambda\left( \eta^2-\eta_0^2\right)^2\,,
\label{45s}
\eeq
see Fig. \ref{eqm}.
This model follows from the discussion above as a phenomenological model.
The parameter $\lambda$ will be fixed shortly, see Eq.~\eqref{lambda_fixed}.
The model in Eq.~\eqref{45s} is quite primitive; however, it captures all qualitative features of the instanton representing the tunneling phenomenon. 

There is one subtle point we must mention. 
All quantities we use in the original 3D theory have dimensions inherent to 3D field theory, while in Eq. (\ref{45s}) we deal with one-dimensional quantum mechanics. 
To avoid the discrepancy, we will measure all relevant quantities in the natural units of mass, i.e., in units of $ev$. 
For simplicity, all numerical factors will be omitted in the remainder of this subsection. 
In these units,
\beq
\eta_0 = v\to e^{-1}\,,\quad M_{\rm vort} = v^2\to e^{-2}, \quad m_\chi \sim m_\phi \sim m = 1\,.
\label{47s}
\eeq
In the topological sector, $M_{\rm vort}$ is the background value of the energy, which is there even at $\tau \to \pm \infty$.
We can subtract this value, much in the same way as was done in \cite{Ievlev:2025ibm}.
The top of the potential hump in (\ref{45s}) corresponds to $M_{\rm sphaleron} =2\pi v^2$; comparing with (\ref{45s}), we see that
\begin{equation}
	\lambda = e^2 \,.
\label{lambda_fixed}
\end{equation}

From the above data, the instanton action in the effective quantum mechanics can be easily obtained.
It is given by
\beq
S_{\rm inst} = m_\chi^3/\lambda = \frac{1}{e^2}\to \frac{m}{e^2} \,,
\eeq
which is dimensionless (as required) and coincides with Eq.~(\ref{43s}).

Thus, we can see that the mixing of the vortices does happen in the bosonic model.
While the model with the potential \eqref{15s} cannot be directly supersymmetrized, one can obtain a close $\mathcal{N}=1$ cousin by deforming the supersymmetric model in Eq.~\eqref{W_almost_bps} as
\begin{equation}
	\Wc = \alpha_1 h\Big(\Phi^\dagger\Phi - v^2\Big) + \alpha_2 h^2 + \alpha_3 h^3 \,.
\label{W_almost_bps_2}
\end{equation}
where $\alpha_{1,2,3}$ are parameters.
However, we do not expect that this supersymmetry would protect the vortex degeneracy.
We postpone the discussion of supersymmetric models to Sec.~\ref{sec:susy_vort_NA}.

\subsection{$\mathbb{CP}(1)$ vortices}

Before passing to the study of non-Abelian superconductors in the next Section, we want to mention another example.
In this setup, non-Abelian degrees of freedom can emerge in a $U(1)$ superconductor.

This example was essentially worked out in \cite{Alonso-Izquierdo:2014cza,Alonso-Izquierdo:2016ieq}.
Here we will briefly review it using an alternative language and give some additional comments with regard to instantons in a topologically non-trivial sector.

Let us start with the so-called $n$-field model, which is often referred to as $O(3)$ or $\mathbb{CP}(1)=SU(2)/U(1)$ in 3D. The target space is a two-dimensional sphere $S_2$ which can be parameterized by a unit vector
\beq
\vec n =\{n^1,\,n^2, \, n^3\}\,,\quad {\vec n}^2 =1\,.
\label{52}
\eeq
The Euclidean action of the $O(3)$ model is given by
\beq
A=\int d^3x \frac{1}{2g^2} \pt_\mu \vec n  \pt^\mu \vec n\,.
\label{53}
\eeq
In fact, this model has two independent real degrees of freedom which can be presented in geometric form as a complex field endowed with the Fubini-Study metric. 
In three dimensions, it has a localized static soliton (a particle) formally coinciding with the Belavin-Polyakov instanton \cite{Polyakov:1975yp}. 
However, this instanton is not what we are interested in.

To get to our main point, let us first deform (\ref{53}) in two ways. 
We introduce the so-called ``real mass'' term
\beq
{\cal L}_m= - m^2 \left(n^3\right)^2,\,\, {\rm in\,\,\, Euclidean}\,\,\,  {\cal L}_m=  m^2\, \left(n^3\right)^2\,.
\label{54}
\eeq
This term explicitly breaks $O(3)$ symmetry down to $U(1)$. 
The energy is minimized at $n^3=0$. 
If we introduce a complex field 
\beq
{\cal  S}=n^1+in^2 \,,
\label{55}
\eeq
then due to the constraint (\ref{52}) the vacuum field configuration takes the form
\begin{equation}
	{\cal S}\bar {\cal S} =|{\cal S}|^2=1\,,\quad \left(n^3\right)_{\rm vac}=0\,,\quad {\cal S}_{\rm vac} = e^{i\alpha} \,.
\label{56}
\end{equation}
Deformation (\ref{54}) is different from the setup of Refs. \cite{Alonso-Izquierdo:2014cza,Alonso-Izquierdo:2016ieq} where certain potentials are used.
The real mass term allows for generalization to $\cal N$=2 supersymmetry.  
Note that the ``standard'' way of adding the twisted mass term \cite{Shifman:2012zz} would change the sign in front of $m^2$ in (\ref{54}); the model then would have just two vacua, the north and the south poles.

The next step is to gauge the $U(1)$ symmetry. 
Corresponding Lagrangian can be written as
%
\beqn
{\cal L}_{\rm gauged} \!\!&=&\!\! -\frac{1} {4e^2}\, F_{\mu\nu}F^{\mu\nu} + \frac{1}{2g^2}\left[\Dc_\mu{\cal S}^\dagger  \Dc^\mu{\cal S}+\pt_\mu n^3\pt^\mu n^3 - m^2 \left(n^3\right)^2\right]\\[2mm]
\Dc_\mu{\cal S}& \equiv& \left(\pt_\mu -A_\mu\right)\cal S\,.\nonumber
\label{57}
\eeqn
The $U(1)$ gauging makes the physical vacuum state unique, i.e. $\alpha$ in (\ref{56}) can be put to zero by a gauge rotation.
The photon eats up the phase of the $\cal S$ field, acquiring the  mass
\begin{equation}
	m_\gamma = \frac{e}{g} \,.
\end{equation}
The mass of the physical Higgs particle is
\begin{equation}
	M_H = \frac{m}{\sqrt 2} \,.
\end{equation}
We assume they are of the same order of magnitude, $m_\gamma \sim M_H$.

Now, let us verify that two distinct static vortices exist in this model (in the quasiclassical approximation $eg \ll 1$).

To build a vortex, we must wind $\cal S$ at large distances from the vortex center, i.e. $ {\cal S} = e^{i\alpha(x)}$ at $r\gg m^{-1}$, where $\alpha$ is now the polar angle, and the radius $r=(x_ix_i)^{1/2}$. 
To guarantee that the energy is finite, at $r\to 0$ we require $|{\cal S}|\to 0$.
Then $\left(n^3\right)^2\to 1$ and $n^3\to \pm 1$.
Thus, the $\mathbb{Z}_2$ symmetry
\begin{equation}
	n^3\to -n^3
\end{equation}
is spontaneously broken in the vortex core, giving rise to two distinct mass-degenerate vortex solutions, much in the same way as in Sec.~\ref{sec:Z2_vort} above.
The vortex mass $M_{\rm vortex}\sim 1/g^2\gg m_\gamma $ in the quasiclassical limit $eg\ll 1$. 

The issue of the interpolating instanton is more subtle. 
We introduce the Euclidean time $\tau$ which varies between $-\infty$ and $+\infty$. 
The instanton is localized in a region around the origin of the Euclidean space, with the characteristic size $\sim m_\gamma^{-1}\times m_\gamma^{-1}$. 
It must interpolate between $n^3=-1$ and $n^3=+1$ vortices in such a way that the corresponding action is finite; otherwise, there is no tunneling. 
The subtlety occurs near the sphaleron trajectory (at $\tau=0$), a situation similar to that discussed in \cite{Ievlev:2025ibm} (see Sec.~4.4.2 there). 

The sphaleron trajectory is as follows,
\beq
\left(n_3\right)^3_{\rm sph} = 0\,,\quad {\cal S} =e^{i\alpha(x)}\,,
\eeq
i.e. ${\cal S}$ winds. 
As a result, $M_{\rm sph}$ is infinite, because the energy density has a singularity $1/r^2$ near the origin, and the integral
\begin{equation}
	\int d^2x\, \frac{1}{r^2}
\end{equation}
is logarithmically divergent.
The sphaleron does not exist on its own. 

However, this singularity is integrable in the instanton action, which is determined by $\int d^3 x$. 
Thus, the action is finite, and the two distinct vortices described above mix through tunneling.

\section{Degenerate vortices in a non-Abelian superconductor}
\label{sec:deg_vort_non-ab}

In this section, we move to non-Abelian models, where vortices with higher multiplicity can be easily constructed.

\subsection{Classical $\mathbb{Z}_N$ vortices in a non-Abelian superconductor }

We start with a model motivated by studies of $\mathcal{N}=2$ supersymmetric QCD (SQCD) in 3+1D; see, e.g., \cite{Shifman:2007ce,Shifman:2009zz} for a review.
For now, we will be discussing only the bosonic version of this model.
Later, we will discuss what happens when we actually add the corresponding fermionic fields and restore supersymmetry.

This is a relativistic field theory in 2+1 dimensions.
However, relativism is actually not important for most of the present discussion, as we will focus on static field configurations.

\subsubsection{Model 1}
\label{sec:model1}

We start with a $U(N)$ gauge theory coupled to scalar matter, 
\begin{multline}
	S=\int d^3 x 
		\Big\{\frac{1}{4 g_2^2}\left(F_{\mu \nu}^a\right)^2+\frac{1}{4 g_1^2}\left(F_{\mu \nu}\right)^2 \\
	 	+ \left|\mathcal{D}_\mu \varphi^A\right|^2+\frac{g_2^2}{2}\left(\bar{\varphi}_A T^a \varphi^A\right)^2+\frac{g_1^2}{8}\left(\left|\varphi^A\right|^2-N \xi\right)^2 \Big\} \,.
\label{action_phi_1}
\end{multline}
Here, $F_{\mu \nu}$ and $F_{\mu \nu}^a$ are the Abelian and the non-Abelian parts of the $U(N)$ gauge field strength.
Generators of $SU(N)$ are normalized as
\begin{equation}
	{\rm tr} T^a T^b = \frac{1}{2} \delta^{ab} \,.
\end{equation}
The matter fields $\varphi^A$ transform in the fundamental representation of the gauge group, and the covariant derivative is given by
\begin{equation}
	\mathcal{D}_\mu=\partial_\mu-\frac{i}{2} A_\mu-i A_\mu^a T^a \,.
\end{equation}
$A = 1, \ldots, N_f$ is the flavor index; we focus on the case $N_f = N$.
The action \eqref{action_phi_1} enjoys $SU(N)_F$ global flavor symmetry.
Each flavor has $N$ color components; denoting the color index as $k = 1, \ldots, N$, we can view $\varphi^A_k$ as an $N \times N$ matrix.

The parameter $\xi$ triggers condensation of $\varphi = \{ \varphi^A_k \}$, 
\begin{equation}
	\expval{ \varphi } = \sqrt{\xi} \begin{pmatrix}
		1 & 0 & \cdots & 0 \\
		0 & 1 & \cdots & 0 \\
		\vdots & \vdots & \ddots & \vdots \\
		0 & 0 & \cdots & 1
	\end{pmatrix}
\end{equation}
This VEV breaks both the color symmetry and the flavor symmetry to a diagonal (color-flavor locked) global subgroup:
\begin{equation}
	U(N)_C \times SU(N)_F \to SU(N)_{C+F} 
\label{locked_group_1}
\end{equation}
The spectrum around this vacuum is gapped, with masses of order $\sim g_{1,2} \sqrt{\xi}$.

Since the $U(1) \in U(N)$ is spontaneously broken by this VEV, this model supports conventional Abrikosov vortices.
Such vortices would correspond to all diagonal components of $\varphi$ winding with the same phase.
However, the most elementary vortices correspond to field configurations where just one component has non-trivial winding at spatial infinity:
\begin{equation}
	\expval{ \varphi }_\text{vor} = \sqrt{\xi} \begin{pmatrix}
		e^{i \alpha } & 0 & \cdots & 0 \\
		0 & 1 & \cdots & 0 \\
		\vdots & \vdots & \ddots & \vdots \\
		0 & 0 & \cdots & 1 
	\end{pmatrix}\,, \quad
	r \to \infty
\label{vortex_winding}
\end{equation}
Here, $(r,\alpha)$ are the polar coordinates on the 2D space.

The winding \eqref{vortex_winding} suggests the following ansatz for the field configuration:
\begin{equation}
	\varphi (r,\alpha)_\text{vor} = \sqrt{\xi} \begin{pmatrix}
		\phi_2(r)^{i \alpha } & 0 & \cdots & 0 \\
		0 & \phi_1(r) & \cdots & 0 \\
		\vdots & \vdots & \ddots & \vdots \\
		0 & 0 & \cdots & \phi_1(r) 
	\end{pmatrix}\,, \quad
	r \to \infty
	\label{vortex_ansatz}
\end{equation}
and a corresponding ansatz for the gauge field.
Such a field configuration further breaks the global symmetry \eqref{locked_group_1} down to $SU(N-1) \times U(1)$.
This suggests that the internal moduli space of such field configurations is
\begin{equation}
	\mathbb{CP}(N-1) = \frac{ SU(N) }{ SU(N-1) \times U(1) }
\label{cpn_1}
\end{equation}
And indeed, one can show this quite explicitly; see \cite{Shifman:2009zz,Shifman:2007ce} for a review.

The mass of this vortex is given by
\begin{equation}
	M_\text{vort} = 2 \pi \xi \,,
\end{equation}
cf. Eq.~\eqref{37s_mass}.
Naturally, \eqref{cpn_1} is supplemented by two translational zero modes, as for the usual Abrikosov vortex.
These modes decouple and will not be important for the considerations here.

\subsubsection{Model 2 (mass deformation)}
\label{sec:model2}

As we will see below, the model \eqref{action_phi_1} (in the purely bosonic case) does not lead to degenerate vortices, because quantum mechanics on the vortex world line \eqref{cpn_1} has a unique ground state.
In order to have non-trivial vortex degeneracy, we introduce a mass deformation of this model.
To further simplify the model, we also take $g_1 = g_2 = g$.

For the deformation, we introduce a complex scalar field $a$ transforming in the adjoint representation of the gauge group $U(N)$.
\begin{multline}
	S=\int d^3 x \Big\{
	\frac{1}{4 g^2}\left(F_{\mu \nu}^a\right)^2 + \frac{1}{g^2} |D_\mu a|^2 + \left|\mathcal{D}_\mu \varphi^A\right|^2 \\
	+\frac{g^2}{2}\left(\bar{\varphi}_A T^a \varphi^A\right)^2 + \frac{g^2}{8}\left(\left|\varphi^A\right|^2-N \xi\right)^2 
	+ \sum_{A=1}^{N} \abs{ (a + M) \varphi }^2
	\Big\} \,,
	\label{action_phi_2}
\end{multline}
Here, $D_\mu$ is a covariant derivative in the adjoint representation, while $M$ is a mass matrix.
We take $M$ to be in a $\mathbb{Z}_N$ symmetric form, with the masses evenly spaced on a circle of some radius $m$:
\begin{equation}
	M = \begin{pmatrix}
		m_1 & 0 & \cdots & 0 \\
		0 & m_2 & \cdots & 0 \\
		\vdots & \vdots & \ddots & \vdots \\
		0 & 0 & \cdots & m_N 
	\end{pmatrix}
	\label{ZN_masses_1}
\end{equation}
\begin{equation}
	m_k = m e^{2 \pi i k / N} \,, \quad
	m > 0 \,.
	\label{ZN_masses_2}
\end{equation}
The last term in \eqref{action_phi_2} forces the adjoint scalar to have a VEV
\begin{equation}
	\expval{ a } = - M \,.
\end{equation}
If the mass matrix $M$ were proportional to an identity matrix, the VEV of $a$ would not change the low-energy theory.
However, in the present discussion the diagonal elements of $M$ are given by \eqref{ZN_masses_2}.
The VEV of $a$ breaks the gauge group to an Abelian subgroup,
\begin{equation}
	U(N) \to U(1)^N \,.
\end{equation}
Thus, the color-flavor locked global symmetry \eqref{locked_group_1} is also broken to $U(1)^{N-1}$.

What happens from the point of view of the vortices \eqref{vortex_winding}?
Strictly speaking, the $\mathbb{CP}(N-1)$ moduli space \eqref{cpn_1} gets lifted.
Only $N$ isolated points remain, corresponding to the $N$ possible vortices where just one particular diagonal component of $\varphi$ is non-trivial.
These correspond to the degenerate vortices that we are interested in.

\subsection{World line action and mixing of vortices }

\begin{figure}[t]
	\centering
	\includegraphics[width=0.5\textwidth]{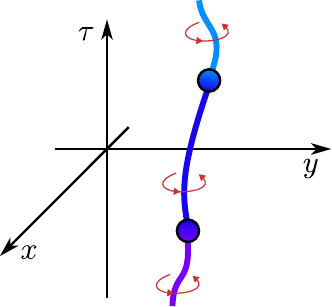}
	\caption{\small
		Churchkhela tunneling process (schematically).
		The vertical axis is the Euclidean time $\tau$.
		The vortex world line has ``lumps'' representing the instantons.
		The line color represents the vortex species.
		Winding around every vortex is the same (shown by red curvy lines with arrows).
	}
	\label{fig:instanton_chain}
\end{figure}

Now let us discuss the low-energy theory on the vortex world sheet in more detail.
Of course, a vortex has two translational zero modes (bosonic), the same as an ordinary Abrikosov vortex.
These modes are decoupled in the small-momentum limit.

On the other hand, the internal moduli space on the vortex gives rise to a non-trivial theory living on the vortex world line.
We start by writing down the low-energy theory corresponding to the model without mass deformation; see \eqref{cpn_1}.
This situation is analogous to the study of the two-dimensional world sheet of non-Abelian strings in 4D SQCD; there, the effective theory comes out to be the $\mathbb{CP}(N-1)$ sigma model \cite{Hanany:2003hp,Auzzi:2003fs,Shifman:2004dr,Hanany:2004ea}; see \cite{Tong:2005un,Eto:2006pg,Shifman:2009zz,Shifman:2007ce,Tong:2008qd} for a review.
We can adapt these results to the present case, where the effective theory lives on a one-dimensional world line.
The (Euclidean) action for this theory is
\begin{equation}
	S_\text{vort} = \int d\tau \Big\{
	\left|(\partial_\tau + i A_\tau) n^k\right|^2 
	+ i\,D\left( \sum_k |n_k|^2 - \beta \right)
	\Big\} \,.
	\label{vortex_theory_1}
\end{equation}
%
Here, $A_\tau$ is an auxiliary gauge field, together with the real scalar $D$ (they play the role of Lagrange multipliers).
The coupling constant $\beta > 0$ is related to the coupling of the bulk 2+1D theory as
\begin{equation}
	\beta \sim \frac{2\pi}{g^2} \,.
\end{equation}
The theory \eqref{vortex_theory_1} has $SU(N)$ global symmetry (there is no $U(1)$ factor because the overall phase of $n_k$ is gauged away).
Naively, the constraint imposed by the last term in \eqref{vortex_theory_1} forces $n_k$ to have a non-zero VEV
\begin{equation}
	\expval{ \sum_k |n_k|^2 } = \beta \,,
\end{equation}
thereby spontaneously breaking this $SU(N)$.
However, we know that continuous global symmetries cannot be spontaneously broken in quantum mechanics.
We stress that the theory \eqref{vortex_theory_1}, while having a non-trivial target space, lives on the one-dimensional world line of the vortex.
This means that the ground state of the effective theory in \eqref{vortex_theory_1} is unique, and there is only one stable species of vortices.

To have a more interesting scenario, let us introduce the mass deformation discussed in Sec.~\ref{sec:model2}, explicitly breaking the global $SU(N)$ down to $\mathbb{Z}_N$.
The effective action reads, in this case,
\begin{equation}
	S_\text{vort} = \int d\tau \Big\{
	\left|(\partial_\tau + i A_\tau) n^k\right|^2 
	+ \sum_k\left|\sigma - m_k\right|^2\, |n^k|^2
	+ i\,D\left( \sum_k |n_k|^2 - \beta \right)
	\Big\} \,.
	\label{vortex_theory_2}
\end{equation}
Here, $m_k$ are the mass parameters from the model \eqref{action_phi_2}, which are given by \eqref{ZN_masses_2}.
The complex scalar $\sigma$ is a new auxiliary field.
Now, the classical minima of the model \eqref{vortex_theory_2} are given by
\begin{equation}
	\expval{ |n_k|^2 } = \beta \, \delta_{k k_0} \,, \quad
	k_0 \in \mathbb{Z}_N \text{ is fixed. }
	\label{vortex_theory_2_vacua}
\end{equation}
Classically, there are $N$ ground states, where only one flavor of $n$'s has a non-zero VEV.
These ground states are permuted by $\mathbb{Z}_N$ symmetry generated by shifts of the index $k$.

On the quantum mechanical level, these ground states may mix.
Let us discuss this question in more detail.

It is known that the 1+1-dimensional version of the theory \eqref{vortex_theory_2} has kinks; see \cite{Dorey:1998yh}.
Each kink interpolates between a pair of neighboring vacua \eqref{vortex_theory_2_vacua}.

In the present case, the model is one-dimensional.
The kinks mentioned above become instantons --- classical solutions localized in Euclidean time $\tau$.
They lead to mixing of the $N$ ground states.

Perturbatively, each ground state $\ket{\psi_k^\text{pert}}$ is localized near the corresponding minimum.
The corresponding low-energy Hamiltonian is simply proportional to an identity matrix, 
\begin{equation}
	H^\text{pert}_\text{eff} \approx E^{(0)}_\text{pert} \mathbb{I}_{N \times N} \,,
\end{equation}
where $E^{(0)}_\text{pert}$ is the perturbative part of the ground state energy in each of the $N$ minima.

In the non-perturbative sector, instantons induce mixing between these states.
This will induce a splitting of a would-be $N$-fold degenerate ground state into a band with a unique\footnote{On the one-instanton level the ground state may turn out to be degenerate in some special cases, but such degeneracy is lifted by multi-instanton effects.} ground state, as we will see shortly.

To understand this splitting quantitatively, let us write down the low-energy Hamiltonian that takes the non-perturbative effects into account.
The first step is computing the instanton actions.
Fortunately, the 0+1D model at hand is closely related to a dimensional reduction of the 1+1D $\mathbb{CP}(N-1)$ model --- or, more precisely, the reduction of the bosonic part of a supersymmetric model.
In that supersymmetric model, the spectrum of kinks is known exactly \cite{Dorey:1998yh}.
As mentioned above, kinks in 1+1D become instantons in 0+1D, and the instanton action coincides with the mass of the ``parent'' kink.
The action of an instanton interpolating from the perturbative vacuum number $i$ to the vacuum number $j$ is given by
\begin{equation}
	S_{ij} = \beta |m_i - m_j| \,,
\end{equation}
where $\beta$ is the coupling, cf. Eq.~\eqref{vortex_theory_1}.
The quantity $S_{ij}$ is smallest when $i = j \pm 1$, while the tunneling amplitudes between more distant states are suppressed, at least for $N > 3$.
This allows us to write down the following low-energy Hamiltonian (leading order, with exponential accuracy):
\begin{equation}
	H_\text{eff} = 
	\begin{pmatrix}
		E^{(0)}	& -\Delta & 0 & \cdots & 0 & -\Delta  \\[4pt]
		-\Delta & E^{(0)} & -\Delta & \ddots & & 0 \\[4pt]
		0 & -\Delta & E^{(0)} & \ddots & \ddots & \vdots \\[4pt]
		\vdots & \ddots & \ddots & \ddots & -\Delta & 0 \\[4pt]
		0 & & \ddots & -\Delta & E^{(0)} & -\Delta \\[4pt]
		-\Delta & 0 & \cdots & 0 & -\Delta & E^{(0)}
	\end{pmatrix}
	\label{H_low_energy_1}
\end{equation}
Here, $E^{(0)}$ is the energy in a particular perturbative ground state given by the matrix element $\bra{\psi_k^\text{pert}} H \ket{\psi_k^\text{pert}}$, while $\Delta$ is the absolute value of the transition amplitude $\bra{\psi_k^\text{pert}} H \ket{\psi_{k+1}^\text{pert}}$, which is the same for any pair of adjacent minima,  
\begin{equation}
	\Delta \sim e^{ - S_{12} } =  e^{ - \beta |m_{k} - m_{k+1}| } \sim  e^{ - 2 \beta m |\sin(  \pi / N )| } \,.
\end{equation}
Minus signs in front of $\Delta$ in \eqref{H_low_energy_1} are not essential and simply reflect a convenient choice%
\footnote{ Strictly speaking, this is true when $N$ is even. For odd $N$, naively a two-fold degeneracy persists on the non-perturbative level, but in this case it will be lifted by further corrections, cf. \cite{Dunne:2020gtk}. Anyway, it is not important for the qualitative picture. } 
of the phases of the perturbative ground states $\ket{\psi_k^\text{pert}}$.

The matrix \eqref{H_low_energy_1} is a real symmetric circulant matrix, whose spectrum is well-known (this is basically the Bloch theorem),
\begin{equation}
	E_k = E^{(0)} - 2 \Delta \cos( \frac{ 2 \pi k }{ N } ) \,, \quad
	k = 1, \ldots, N \,.
	\label{low_energy_spectrum_1}
\end{equation}

In either case, the ground state wave function is a symmetric combination of our basis $\ket{\psi_k^\text{pert}}$, with the ground state energy 
\begin{equation}
	E_\text{ground} \approx E^{(0)} - 2 \Delta \sim E^{(0)} - 2 m e^{ - 2 \beta m |\sin(  \pi / N )| } \,.
\end{equation}
This formula gives a good approximation at least when $N$ is large enough.
When $N$ is small, the approximate form \eqref{H_low_energy_1} does not work that well (the target space of $\mathbb{CP}(N-1)$ relevant at low energies becomes very different from a circle).
Still, a more careful analysis shows that the ground state is again unique \cite{Fujimori:2016ljw}.

From the point of view of the 2+1D bulk, these world line instantons are nothing but monopole-instantons.
These monopole-instantons cannot exist in the bulk by themselves --- instead, they are confined to the vortex world lines; see also \cite{Shifman:2004dr}. 
The world line then resembles a vine or a churchkhela sweet; see Fig.~\ref{fig:instanton_chain}.

Let us reiterate what we observe here.
We have a superconducting medium with vortices which classically appear to have $N$-fold degeneracy with respect to some internal quantum numbers.
However, because of the world line instantons, this degeneracy is lifted, and the lowest-energy vortex is unique.

\subsection{Supersymmetric case}
\label{sec:susy_vort_NA}

Having discussed the vortices in bosonic theories, let us now add supersymmetry and see what changes.

\subsubsection{Eight supercharges in the 3D bulk ($\mathcal{N}=4$)}

Now, let us ask the question: what happens if we consider a supersymmetric version of the setup above?
It often happens that, in the presence of fermions, the instantons acquire fermionic zero modes, which suppress the tunneling amplitudes between perturbative ground states.

As mentioned above, the model in Eq.~\eqref{action_phi_2} is motivated by the bosonic part of the 3+1D SQCD with $\mathcal{N}=2$ supersymmetry (eight supercharges).
Basically, one can arrive at the action \eqref{action_phi_2} by a dimensional reduction from 3+1D and setting the holonomy of the gauge field to zero (it is not turned on in the vortex background that we are interested in).

Let us now consider a reduction of the full supersymmetric SQCD by compactifying it on a circle.
After the reduction, we will arrive at 2+1D SQCD with $\mathcal{N}=4$ supersymmetry (again, eight supercharges).
The vortices discussed above become 1/2 BPS objects.
The world line $\mathbb{CP}(N-1)$ quantum mechanics \eqref{vortex_theory_2} becomes supersymmetric with four supercharges.
In the context of the dimensional reduction, this QM can be viewed as descending from the world sheet effective $\mathbb{CP}(N-1)$ theory on the non-Abelian vortex in 3+1D SQCD; the latter has $\mathcal{N}=(2,2)$ supersymmetry.

Since the vortex configuration is determined first of all by its bosonic profiles, and these do not change when we include fermions, classically we still have degenerate vortices exchanged by the internal $\mathbb{Z}_N$ symmetry.
In the purely bosonic case, these vortices mix due to world line instantons.
Let us see what happens in the supersymmetrized version of the model.

Consider the supersymmetric quantum mechanics on the vortex world line.
In principle, supersymmetry does not automatically guarantee that all classical ground states persist at the quantum level.
The vacua with different fermionic structure do not mix in supersymmetric QM, but vacua with the same fermionic structure do mix.
For example, this happens in a theory with one complex supercharge and a quartic superpotential; see, e.g., \cite{tongQM}.
In general, the surviving degeneracy of ground states in quantum theory cannot exceed the number of linearly independent structures that one can form from the fermionic degrees of freedom at hand.

The $\mathbb{CP}(N-1)$ QM with four real supercharges (two complex) has $4 (N-1)$ fermionic degrees of freedom, which can be enough to save all $N$ ground states.
In this case, the ground states are in one-to-one correspondence with the space of harmonic forms on $\mathbb{CP}(N-1)$ \cite{Witten:1982im} (see also \cite{Dunne:2012ae,tongQM}), and the number of ground states is given by the Witten index, which coincides with the Euler characteristic of the target space:
\begin{equation}
	\#_\text{ground} = \chi ( \mathbb{CP}(N-1) ) = N \,.
	\label{cpn_susy_counting}
\end{equation}
All ground states are described by even-degree forms.
The instanton has exactly one fermionic zero mode, and because of that all transition amplitudes vanish.

The above argument for $\mathbb{CP}(N-1)$ QM and the ground state counting \eqref{cpn_susy_counting} turns out to be correct not only in the mass-deformed model, but also persists in the limit $m \to 0$.

We can conclude that when the 2+1D theory has eight supercharges, the BPS vortices remain degenerate at the quantum level.

\subsubsection{Less supersymmetry ($\mathcal{N}=2$ or $\mathcal{N}=1$ in 2+1D)}

What happens if we reduce the number of supercharges in the bulk?
Depending on the details, different scenarios are possible.

One can break $\mathcal{N}=4$ to $\mathcal{N}=2$ by deforming the bulk 2+1D theory.
This can be achieved, e.g., by deforming the 3+1D SQCD and then dimensionally reducing to 2+1D.
We will mention here two well-studied examples, where the supersymmetry is softly broken by a mass term for the adjoint supermultiplet.

\vspace{10pt}

In the first example \cite{Shifman:2008wv}, 
the deformation in 3+1D keeps two supercharges on the flux tube world sheet.
Two fermionic zero modes from the translational sector mix with internal $\mathbb{CP}(N-1)$ moduli, breaking the other two of the four supercharges from the undeformed case.
The supersymmetry on the world sheet turns out to be spontaneously broken, and, consequently, the Witten index is zero.
However, the ground state degeneracy remains $N$, as the energies of all vacua become shifted from zero by the same amount.
These ground states are associated with spontaneous breaking of the chiral symmetry on the world sheet, $\mathbb{Z}_{2N} \to \mathbb{Z}_2$.

When we perform a dimensional reduction, the flux tubes with 1+1D world sheets become vortices with 0+1D world lines.
Chiral symmetry disappears.
Therefore, we expect that the ground state degeneracy is not protected and becomes lifted.

In the second example \cite{Ievlev:2017atl}, 
the deformation just eliminates the world sheet supersymmetry completely.
There are no massless fermions in the $\mathbb{CP}(N-1)$ internal sector.
Therefore, the ground state degeneracy is lifted from the start.

Moving further, consider the case where the supersymmetry in the 2+1D theory reduces to $\mathcal{N}=1$ (two supercharges).
In this situation, the vortex cannot be BPS.
This can be seen from the fact that in 2+1D, a vortex with winding number one always has two translational zero modes, and therefore it can be 1/2 BPS only in a theory with four (or more) supercharges; see also the discussion in Sec.~II.A of \cite{Ritz:2004mp} or in Sec.~2.2.2 of \cite{Shifman:2009zz}.
In this case, while it may be possible that the instanton mixing is still suppressed (cf. the discussion in Sec.~3.8 of \cite{Ievlev:2025ibm}), generically we cannot expect that.
Rather, the most natural assumption is that the degeneracy is lifted in this case as well.

\vspace{10pt}

To reiterate, in the example of a non-Abelian superconductor considered here, we see degenerate vortices on the classical level.
However, this degeneracy persists on the quantum level only in the case with $\mathcal{N}=4$ supersymmetry in 2+1D (eight supercharges), which gives four supercharges on the vortex.

When the amount of supersymmetry in the bulk is lowered, the degenerate vortices mix due to world line instantons.
As a result, the degeneracy is lifted, and there is a single lowest-energy vortex.

We expect that it should be possible to have an $\mathcal{N}=2$ model where the Witten index of the world line quantum mechanics does not vanish (cf. \cite{Bullimore:2019qnt}) and the vortices stay degenerate.
We leave detailed investigation of this possibility to future work.

\section{Conclusions}
\label{sec:concl}

In this paper, we extended the study of quantum lifting of classical soliton multiplicity from 1+1 dimensions to genuinely 2+1-dimensional settings, focusing on vortices. 
The central message is that whenever the classical vortex sector contains several distinct species with identical mass and topological charge, the effective 0+1D dynamics on the vortex world line can typically admit Euclidean-time instantons that interpolate between these would-be distinct states, splitting the classical degeneracy. 

We illustrated this mechanism first in Abelian superconductors. 
In the $U(1)$ model supplemented by a neutral scalar with an unbroken $\mathbb{Z}_2$ symmetry in the vacuum, the vortex core spontaneously breaks this $\mathbb{Z}_2$ and supports two classically degenerate vortices.
These states can tunnel into each other through a finite-action world-line instanton, so the corresponding vortex ``doublet'' is not protected in the purely bosonic theory. 
A simple effective quantum-mechanical description captures the parametric behavior of the tunneling action, $S_\text{inst} \sim m/e^2$, with the splitting controlled at weak coupling. 

We then analyzed non-Abelian superconductors, where internal moduli are naturally present. 
In the $U(N)$ setup, the undeformed vortex has a $\mathbb{CP}(N-1)$ internal moduli space, while an appropriate mass deformation lifts the moduli space to $N$ isolated classical vortices, providing an explicit realization of an $N$-fold classical multiplicity. 
The resulting world-line theory admits instantons (descended from $\mathbb{CP}(N-1)$ kinks), which generate mixing and split the $N$ classical vacua into a band with a unique lowest-energy state in the generic bosonic case. 
From the bulk viewpoint, these world-line instantons can be understood as monopole-instantons confined to the vortex.
Supersymmetry, if present, can still protect the vortex degeneracy.

\vspace{5pt}

Logically, the next step would be a discussion of degenerate monopoles in a 3+1D setting, where they become particle-like states.
One can generalize the basic idea of Sec.~\ref{sec:deg_vort_abelian}.
Namely, the simplest way is to introduce an extra field that vanishes in the vacuum, but which has a non-zero VEV at the monopole core.
If there is a discrete symmetry associated with this extra field, then that symmetry is spontaneously broken by the monopole, and one can obtain multiple monopole species, degenerate in energy.

Another potentially interesting avenue could be an investigation of 2D theories with $\mathcal{N}=(0,2)$ supersymmetry.
An example of this is a heterotic $\mathbb{CP}(1)$ with ``matter'' (e.g. a $\mathbb{CP}(1) \times \mathbb{C}$ model).
It may be possible to construct degenerate kinks that are BPS and hence conserve one supercharge.
In this case, it would be interesting to understand the structure of the fermion zero modes on the kink, and also what happens to the world line instantons and tunneling in this case.

We leave these questions for future work.

\section*{Acknowledgments}

We thank Jarah Evslin for helpful discussions. This work is supported in part by U.S. Department of Energy Grant No. de-sc0011842.

\appendix

\section{\boldmath$\mathcal{N}=1$ supersymmetry in three dimensions  }
\label{sec:3d_susy}

In this Appendix, we are going to review some basic constructions used in 3D theories with $\mathcal{N}=1$ supersymmetry (two supercharges).
For the most part, we are going to follow \cite{Gates:1983nr}, although our notation is somewhat modernized to be consistent with the generally accepted 4D conventions; see, e.g., \cite{McKeon:2001su} and Secs. 10.2-10.6 of \cite{Shifman:2012zz}.

\subsection{Spinors in 3D}

The superspace is parameterized by coordinates $x^\mu$ and $\theta^\alpha$, where
$$x^\mu =\{x^1,\,x^2,\,x^3\}\equiv \{t,x,z\} \,, $$
and 
the fermion directions of superspace are parametrized by two real Grassmann numbers $\theta^1$ and $\theta^2$ forming a Majorana spinor in 3D,
\beq
\theta^\alpha = \left(
\begin{array}{c} \theta^1 \\
	\theta^2
\end{array}
\right)\!.
\eeq
We use Greek letters from the beginning of the Greek alphabet ($\alpha ,\, \beta , \ldots$) as spinorial indices for 3D Majorana spinors, while those from the end of the alphabet ($\mu,\, \nu , \ldots$) are used as vectorial indices.

Our convention for the metric tensor is (note that it is different from the one used in \cite{Gates:1983nr})
\begin{equation}
	g^{\mu\nu}={\rm diag} \{1,-1,-1\} \,.
\label{3d_metric}
\end{equation}
The gamma matrices satisfy the standard algebra
\beq
\gamma^\mu\,\,\, {\rm for} \,\,\,\mu = 0,1,3\,\,\,{\rm and}\,\,\, \{\gamma^\mu\gamma^\nu \} = 2 g^{\mu\nu}\,.
\label{3s}
\eeq
It is convenient to choose the Majorana representation where the $\gamma$ matrices are purely imaginary, and $\gamma^0$ is Hermitian while $\gamma^1$ and $\gamma^3$ are anti-Hermitian.
Explicitly,
\beq
(\gamma^0)_{\alpha\beta} = (\sigma_2)_{\alpha\beta},\,\,\,\,\,\,  (\gamma^1)_{\alpha\beta} = i(\sigma_1)_{\alpha\beta}, \,\,\,\,\,\,  (\gamma^3)_{\alpha\beta} =  -i(\sigma_3)_{\alpha\beta}\ .
\label{2spp}
\eeq
Dirac conjugation is defined as
\begin{equation}
	\bar\psi= \psi^\dagger \gamma^0 = ( i \psi_2, - i \psi_1 )	 \,.
\end{equation}

Raising and lowering of spinorial indices is performed as usual -- by multiplying a spinorial quantity from the  left by 
\beq
\varepsilon^{\alpha\beta} = i(\sigma_2)_{\alpha\beta} =\left( \begin{array}{cc} 0&1 \\-1&0
\end{array}
\right),\,\,\,\,
\varepsilon_{\alpha\beta} = -i(\sigma_2)_{\alpha\beta} =\left( \begin{array}{cc} 0&-1 \\1&0
\end{array}
\right)\,.
\eeq
With this convention,
\begin{equation}
	\varepsilon^{\alpha\beta} \varepsilon_{\beta\delta}= \delta^\alpha_\delta \,.
\end{equation}

In 4D supersymmetric theories, along with the gamma matrices it is very convenient to use 
$\sigma^\mu$ matrices, namely  (see e.g. \cite{Shifman:2012zz}) 
$(\sigma^\mu )_{\alpha\dot\beta} =\{1, \sigma_1, \sigma_2, \sigma_3\}_{\alpha\beta}$.
In three dimensions we can define the analogous tensor as follows,
\begin{equation}
\begin{aligned}
	(\Gamma^\mu)_{\alpha\beta} &= (\gamma^0\gamma^\mu)_{\alpha\beta}  =\{1, \sigma_3, \, \sigma_1\}_{\alpha\beta} \,, \\
	(\Gamma^\mu)^{\alpha\beta} &= \varepsilon^{\alpha\delta}\varepsilon^{\beta\rho} (\Gamma^\mu)_{\delta\rho}= \{1, -\sigma_3, \, -\sigma_1\}_{\alpha\beta} \,.
\end{aligned}
\end{equation}
All $\Gamma^\mu$ matrices defined this way are Hermitian, real and symmetric. 
With their help we can convert 3D vectors into the spinor representation, where they become {\em symmetric} $2\times 2$ matrices.
For example, a vector $A_\mu = \{A^0, -\vec A \}$ becomes
\beq
A_{\alpha\beta} \equiv A_\mu (\Gamma^\mu)_{\alpha\beta} = (A_0-A^1\sigma_1-A^3\sigma_3)_{\alpha\beta}.
\eeq
while the inverse transformation is 
\begin{equation}
	A^\mu = \frac{1}{2} A_{\alpha\beta} (\Gamma^\mu)^{\alpha\beta} \,.
\end{equation}
Other useful identities involving vectors:
\begin{equation}
\begin{aligned}
	A_\mu B^\mu &= A_0B_0 -\vec A\vec B= \frac 1 2 A_{\alpha\beta} B^{\beta\alpha} \,; \\
	\partial_{\alpha\beta}\partial^{\gamma\beta}   &= \delta^\gamma_\alpha\,\partial^\mu\partial_\mu \,.
\end{aligned}
\end{equation}
Other useful spinorial identities:
\begin{equation}
\begin{aligned}
	&(\Gamma^\mu)_{\alpha\beta} (\Gamma^\nu)^{\alpha\beta} = 2 g^{\mu\nu} \,; \\
	&\theta^\alpha\theta^\beta =-\frac 1 2 \varepsilon^{\alpha\beta}\big(\theta^1 \theta_1 +\theta^2 \theta_2 \big)\equiv -\frac 1 2 \varepsilon^{\alpha\beta}\theta^{\{2\}} \,; \\
	&\theta^{\{2\}} \equiv \big(\theta^1 \theta_1 +\theta^2 \theta_2 \big)= -2\theta^1 \theta^2 = i \bar{\theta} \theta \,; \\
	&\theta^\alpha{\theta}_\alpha^\prime = {\theta}^{\prime\,\beta} \theta_\beta \,; \\
	&\frac{\pt}{\pt\theta^\alpha} \, \frac{i}{2} \theta^{\{2\}} = i\varepsilon_{\alpha\beta}\theta^\beta \,.
\end{aligned}
\label{useful_spinors}
\end{equation}

Note that in three dimensions with one real superfield, the measure $d\theta^1 d\theta^2$ would be anti-Hermitian\footnote{Conjugation switches the order of the terms in the product, and bringing it back to $d\theta^1 d\theta^2$ yields an extra minus sign. }. 
To make it Hermitian in the super-action, we will replace it by  
\beq
i \, d\theta^1 d\theta^2 \stackrel{ \equiv\,\rm def}{\longrightarrow} d^2\theta \,.
\label{aa10}
\eeq
Then we will normalize  the Berezin integral as follows,
\beq
\int d\theta^1 d\theta^2 \,\theta^1\theta^2 = \frac{i}{2} \int d^2\theta \, \theta^{\{2\}} = 1\,.
\eeq

\subsection{Superspace and superfields}

There is no chirality in the minimal 3D theory. 
The superspace is based on five coordinates --  $c$-numerical $ (t, x^1, x^3)$ and two Grassmann $\theta^{1,2}$.
Supercoordinate transformations are defined as
\beqn
{\mathcal M} &=& \{x^\mu, \theta^\alpha\} \stackrel{\rm susy}{\longrightarrow} \{x^\mu +\delta x^\mu,\,\theta^\alpha+\delta\theta^\alpha \}, 
\nonumber\\ [2mm]
\delta\theta^\alpha &=& \epsilon^\alpha\,,\quad \delta x_{\alpha\beta} = -i ( \theta_\alpha\epsilon_\beta + \theta_\beta\epsilon_\alpha),
\nonumber\\ [1mm]
x_{\alpha\beta} &=& x_\mu (\Gamma^\mu)_{\alpha\beta},\quad x^\mu =\frac 1 2 x_{\alpha\beta} \left(\Gamma^\mu\right)^{\alpha\beta},
\eeqn
with fermionic parameters  $\epsilon_\alpha$. It is easy to see that under the above transformations the components of the {\em real} scalar superfield 
\beq
\Phi_{\rm r} = \varphi + i\theta^\alpha \psi_\alpha + \frac i 2 \theta^{\{2\}} F\,,
\label{20sp}
\eeq
transform through each other as follows,
\beq
\delta\varphi =\epsilon^\alpha \psi_\alpha, \quad \delta\psi_\alpha =-i\epsilon^\beta\, \partial_{\alpha\beta}\,\varphi +\epsilon_\alpha F,\quad \delta F =-i\epsilon^\beta\,\partial_{\alpha\beta}\psi^\alpha\,.
\eeq
Here we omit the ``flavor'' supersript $a$ introduced in (\ref{sigmaLp}).

\vspace{2mm}

Spin-covariant derivatives on the superspace are defined as
\beq
\label{16s}
D_\alpha =\left(i\frac{\partial}{\partial\theta^\alpha}  + \theta^\beta\partial_{\alpha\beta}\right),\qquad
\partial_{\alpha\beta}= \frac{\partial}{\partial x^\mu} (\Gamma^\mu)_{\alpha\beta} \,.
\eeq
With this definition at hand,
\begin{equation}
	D_\alpha \Phi_{\rm r} = -\psi_\alpha -\theta_\alpha F+\theta^\beta\left(\pt_{\alpha\beta}\varphi\right) -\frac i 2\, \theta^{\{2\}}\pt_{\alpha\beta} \psi^\beta \,.
\end{equation}
Other useful expressions:
\begin{equation}
	D_\alpha D_\beta=i\partial_{\alpha\beta}-\varepsilon_{\alpha\beta}D^{\{2\}} \,, \quad
	\Big\{D_\alpha D_\beta\Big\} = 2 i \partial_{\alpha\beta} \,.
\end{equation}

\subsection{Wess-Zumino type model}

Now we have all we need to write down the kinetic part of the action for a real matter field in superspace,
\beqn
S_{\rm r}&=&\int d^3 x\, d^2\theta \frac{i}{4}  \left(D^\alpha\Phi_{\rm r}\right)^\dagger (D_\alpha\Phi_{\rm r})\nonumber\\[2mm]
&=&\int d^3 x\, \frac{1}{2}\left[\pt_\mu\varphi\pt^\mu\varphi +\psi^\alpha i\pt_{\alpha\beta}
\psi^\beta + F^2
\right]\,,
\eeqn
see also our definition for the $\theta$-measure (\ref{aa10}).
The potential interaction is usually introduced through a superpotential. 
Because of the absence of the chiral subspace in 3D, we cannot do it in the standard way. 
Nevertheless, if we introduce a function ${\cal W} (\Phi_{\rm r})$ and add it in the Lagrangian as
\beq
{\cal L}_{\cal W}= \int d^2\theta  {\cal W} (\Phi_{\rm r}) \,,
\label{a20}
\eeq
some general features of the superpotential survive.
For instance, the mass term is given by
\begin{equation}
	{\cal W} (\Phi_{\rm r}) = \frac{1}{2} m \Phi_{\rm r}^2 \,, \quad
	\mathcal{L}_\text{mass} = \frac 1 2 m \bar\psi\psi +m\varphi F \,.
\end{equation}
Note that in the present case with Majorana fermions $\bar\psi\psi = 2 i \psi_2 \psi_1$, cf. Eq.~\eqref{useful_spinors}.
In the case of a more general superpotential, the scalar potential and the fermion mass term are given by
\begin{equation}
	U(\varphi) = \frac{1}{2} \left( \pdv{ \mathcal{W} }{ \Phi_{\rm r} } \right)^2 \Bigg|_{ \theta = 0 } \,; \quad
	\mathcal{L}_{\psi \text{mass}} = \frac{1}{2} \left( \pdv[2]{ \mathcal{W} }{ \Phi_{\rm r} } \right) \Bigg|_{ \theta = 0 }
\end{equation}
where we have eliminated the auxiliary field $F$ by virtue of its equation of motion.

\subsection{Complexification and gauging}

In order to have matter that is charged under a $U(1)$ gauge group, we, quite plainly, need a $U(1)$ symmetry in the matter sector.
To this end, we consider the above setup with two real scalar superfields $\Phi_{\rm r}^a$, $a=1,2$ defined in (\ref{20sp}), which we combine into one complex superfield 
\beq
\Phi_{\rm c} 
= \frac{1}{\sqrt 2} (\Phi_{\rm r}^1+i \Phi_{\rm r}^2)
= \phi + i\theta^\alpha \Psi_\alpha + \frac i 2 \theta^{\{2\}} \Fc \,,
\label{20s}
\eeq
same as was defined in Eq.~(\ref{complex}).
In what follows, we endow the complex superfield (\ref{20s}) with unit electric charge.

The gauge superfield  $V_\alpha$  can be constructed following the line of reasoning of Gates et al. \cite{Gates:1983nr}.
Recall that in the non-supersymmetric QED the photon field is transformed under the gauge transformation as
\beq
(A_\mu)_{\rm gt} = A_\mu+\partial_\mu \alpha (t,x) \,;
\label{17s}
\eeq
here, $x = (x^1,x^2)$ are the spatial coordinates, and the subscript ``gt'' stands for gauge-transformed.
This transformation can be derived, for instance, from the requirement
\beq
\big({\mathcal D_\mu} \phi(x)\big)_{\rm gt} = e^{i\alpha(x)} \big({\mathcal D_\mu} \phi(x)\big)\,\,\, {\rm if}\,\,\, \phi_{\rm gt}(x) = e^{i\alpha (x)}\phi(x) \,,
\label{18s}
\eeq
where $\phi$ is a matter field and  ${\mathcal D_\mu}$ is the gauge covariant derivative,
\beq
{\mathcal D_\mu} = \partial_\mu -iA_\mu .
\eeq
Next, we do the same for supergauge transformations. 
In analogy with the second term in (\ref{18s}), we have
\beq
\Phi_{\rm gt} = e^{i K} \Phi \,,
\label{22s}
\eeq
where $K$ is a real scalar superfield,
\beq
K =\kappa +i \theta^\alpha\xi_\alpha + \frac i 2 \theta^{\{2\}}\chi \,.
\label{24s}
\eeq
The gauge covariantization starts from the spinorial derivative $D_\alpha$ \cite{Gates:1983nr}, 
\beq
D_\alpha \longrightarrow \mathcal{D}_\alpha = \mathcal{D}_\alpha - i V_\alpha \,.
\label{25s}
\eeq
In analogy with \eqref{18s}, we demand that
\beq
\big(\mathcal{D}_\alpha \Phi\big)_{\rm gt} = e^{i K} \big(\mathcal{D}_\alpha \Phi  \big) \,.
\label{gt_susy_Dphi}
\eeq
This can be achieved provided that
\beq
\big(\mathcal{D}_\alpha \big)_{\rm gt} = e^{i K}\mathcal{D}_\alpha e^{-i K}\,,
\eeq
which in turn implies
\beq
(V_\alpha)_{\rm gt} =V_\alpha + \delta V_\alpha\,, \quad \delta V_\alpha = D_\alpha K\,.
\label{lish}
\eeq
Equation (\ref{lish}) prompts the form of the gauge connection $V_\alpha$.
Indeed, from Eqs. (\ref{16s}) and (\ref{24s}) we conclude that
\beq
D_\alpha K = - \xi_\alpha+ \theta^\beta\partial_{\alpha\beta} \kappa -\theta_\alpha\chi -\frac i 2 \theta^{\{2\}}\partial_{\alpha\beta}\xi^\beta\,.
\label{27sp}
\eeq
In components, $V_\alpha$ takes the form
\beq
V_\alpha = \zeta_\alpha +\theta^\beta A_{\alpha \beta}+\theta_\alpha\sigma + \frac i 2 \theta^{\{2\}}\left(\lambda_\alpha +\partial_{\alpha\beta}\zeta^\beta \right).
\label{28s}
\eeq
Comparing (\ref{27sp}) and (\ref{28s}), we see that we can choose a gauge which is analogous to the Wess-Zumino gauge in 4D:
\beq
(V_\alpha)_{\rm WZ} = \theta^\beta A_{\alpha \beta}+ \frac i 2 \theta^{\{2\}}\lambda_\alpha\,.
\label{29s}
\eeq
In what follows, we will omit the subscript WZ.

A superfield representing the gauge field strength can be defined as
\beq
W_\alpha =\frac 1 2 D^\beta D_\alpha V_\beta =\lambda_\alpha +\tilde{F}_\mu \big( \Gamma^\mu \big)_{\alpha \beta}\,\theta^\beta +
i \theta^{\{2\}}  \partial_{\alpha\beta} \lambda^\beta \,,
\label{30s}
\eeq
where $\tilde{F}_\mu$ is the dual photon field tensor,
\beq
\tilde{F}_\mu =\frac 1 2 \varepsilon_{\mu\nu\rho}F^{\nu\rho} = \varepsilon_{\mu\nu\rho}\, \partial^\nu A^\rho\,.
\label{dual_field_strength}
\eeq
The Super-Bianchi identity has the form
\begin{equation}
	D^\alpha W_\alpha =0 \,.
\end{equation}
This formula can be trivially verified if we use the component expansion (\ref{30s}), the property $\partial^\mu \tilde{F}_\mu\equiv 0$ following from \eqref{dual_field_strength}, and the fact that all three $\Gamma$ matrices are symmetric with respect to the spinorial indices.

\subsubsection{Three-dimensional  $\mathcal N$=1 Supersymmetric QED}

In three dimensions, the photon field $A_\mu$ represents one physical degree of freedom, and so does the Majorana two-component spinor $\lambda_\alpha$ --- the lowest component in the superfield (\ref{30s}). 
The photon/photino kinetic terms in the action take the form
\begin{equation}
	\begin{aligned}
		S_{\gamma\, \rm kin } &= \int d^3 x\,d^2\theta \,\frac{1}{2e^2} W^\alpha W_\alpha \,, \\
		\Lc_{\gamma\, \rm kin} &= \frac{1}{2 e^2} \Big(- {\tilde F}_\mu {\tilde F}^\mu + i\lambda^\alpha \partial_{\alpha\beta}\lambda^\beta\Big) \,.
	\end{aligned}
	\label{33s}
\end{equation}
The kinetic part of the matter action can be written down covariantly as
\beq
S_{\rm kin\,  matter} =  \int d^3 x\,d^2\theta \, \left(\mathcal{D}^\alpha \Phi_{\rm c}^\dagger \mathcal{D}_\alpha\Phi_{\rm c}\right) \,,
\label{35s}
\eeq
where the covariantized spinor derivative $\mathcal{D}_\alpha\Phi_{\rm c}$ is defined above Eqs. (\ref{25s}) and (\ref{29s}).
The ``superpotential'' term has the standard form%
\footnote{Similar to the discussion of the real superfields above Eq.~\eqref{a20}, in ${\mathcal N}=1$ SQED no {\em bona fide} chiral superpotential exists because there are no chiral sub-spaces. This is the reason for the quotation marks. The ``superpotential'' in (\ref{36s}) depends on both $\Phi_{\rm c}$ and $\Phi_{\rm c}^\dagger$. Nevertheless, many conventional relations are still valid. }
\beq
S_{\rm matter \,{\mathcal W}} =  \int d^3 x\,d^2\theta \, {\mathcal W(\Phi_{\rm c} , \Phi_{\rm c}^\dagger)}\,.
\label{36s}
\eeq
The $\theta^{\{2\}}$ component of the superfield $\Phi_{\rm c}$ from Eq.~\eqref{20s} (i.e. the $\Fc$ term) enters the action without derivatives and, therefore, can be eliminated by using the classical equation of motion.
In this way, we derive the scalar potential $U$,
\begin{equation}
	U(\phi, \bar{\phi}) =\left. \frac{\partial {\mathcal W}}{\partial \Phi_{\rm c}} \, \frac{\partial {\mathcal W}}{\partial \Phi_{\rm c}^\dagger}\right|_{\theta=0} \,.
\end{equation}

The ``superpotential'' ${\mathcal W}$ which will be responsible for Higgsing of the gauge field and its superpartner $\lambda$ is chosen as follows. 
Usually, to this end, we combine quadratic and cubic matter superfields, but in the case at hand this is impossible. 
Indeed, no cubic gauge-invariant term exists in the case at hand. 
Therefore, we are forced to invoke a quartic term, 
\beq
{\mathcal W} = m \Phi_{\rm c}^\dagger \Phi_{\rm c} - \frac{g}{2} (\Phi_{\rm c}^\dagger \Phi_{\rm c} )^2\,,
\label{quartsup}
\eeq
where $m$ and $g$ are assumed to be real positive parameters. 
This yields
\begin{equation}
	U(\phi, \bar{\phi}) = (\phi\,\phi^\dagger)\left[m-g (\phi\,\phi^\dagger)\right]^2 \,.
\end{equation}

Since in the case at hand the mass dimension of the superpotential is 2, and that of the superfield
$\Phi_{\rm c}$ is $1/2$ , $m$ has mass dimension 1 while $g$ is dimensionless.
The above expression implies that the scalar potential will have a term of sixth order in the matter field. 
This does not spoil renormalizability of our 3D theory. Moreover, in addition to the Higgs regime $\Phi_{\rm vac} \neq 0$ we will also have a $\Phi_{\rm vac} = 0$ vacuum with the un-Higgsed gauge field.

\subsection{Central (brane) charges}
\label{sec:DL_charge}

The \none SUSY algebra in three dimensions $\{t,x,z\}$ is as follows:
\beqn
\{ Q_\alpha\,,  { Q}_\beta \}&=&(\Gamma^\mu)_{\alpha\beta}\left( 
P_\mu +\Zc_\mu\right),\quad
\left[Q_\alpha, P_\mu\right]=\left[Q_\alpha, {\cal Z}_\mu\right]=0\,,\\[1mm]
\mu,\,\nu
&=&0,1,3,\nonumber
\label{alg}
\eeqn
where $P_\mu $, ($\mu
=0,1,3$) is the energy-momentum vector.

If we introduce a topologically (i.e. nondynamically) conserved two-index tensor
\beq
\zeta_{\mu\nu} =\varepsilon_{\mu\nu\rho}\pt^\rho \Wc \,,
\eeq
we find a domain line (wall) charge defined as
\beq
\Zc_\mu =\int d^2 x\, \zeta_{\mu 0}\,.
\eeq
For a static domain line stretched along the $x$ axis (i.e. its profile depends only on $z$),  $\Zc_\mu$ reduces to
\beq
L\, {\cal Z}_{\mu=1}=\int  {\rm d} z\, \partial_z \phi^a 
\,\partial_a {\cal W}\,,\quad  \quad {\cal Z}_{\mu=0\, {\rm or}\, 3}=0\,,
\label{cch}
\eeq
and $P_\mu \Zc^\mu =0$ as is required.

We can conclude that, while $\mathcal{N}=1$ superalgebra in 3D admits codimension-1 BPS objects, there is no central charge for a codimension-2 state (like a vortex).

\bibliographystyle{JHEP}

\bibliography{vortices}

\end{document}